\documentclass[12pt,a4paper]{article}
\usepackage[utf8]{inputenc}
\usepackage{graphicx}%
\usepackage{multirow}%
\usepackage{amsmath,amssymb,amsfonts}%
\usepackage{amsthm}%
\usepackage{mathrsfs}%
\usepackage[title]{appendix}%
\usepackage{xcolor}%
\usepackage{textcomp}%
\usepackage{manyfoot}%
\usepackage{booktabs}%
\usepackage{algorithm}%
\usepackage{algorithmicx}%
\usepackage{algpseudocode}%
\usepackage{listings}%
\usepackage{natbib}
\usepackage{tikz}
\usetikzlibrary {shapes.geometric}
\usepackage{makecell}
\usepackage{anyfontsize}
\usepackage{parskip}
\usepackage{subfig}

\definecolor{sangria}{rgb}{0.57, 0.0, 0.04}
\newcommand{\blueline}{\raisebox{2pt}{\tikz{\draw[-,blue,solid,line width = 0.9pt](0,0) --(5mm,0);}}}

\newcommand{\blackline}{\raisebox{2pt}{\tikz{\draw[-,black,solid,line width = 0.9pt](0,0) --(5mm,0);}}}
\newcommand{\orangeline}{\raisebox{2pt}{\tikz{\draw[-,orange,solid,line width = 0.9pt](0,0) --(5mm,0);}}}

\newcommand{\redline}{\raisebox{2pt}{\tikz{\draw[-,red,solid,line width = 0.9pt](0,0) -- (5mm,0);}}}

\newcommand{\greensquare}{\raisebox{0pt}{\tikz{\filldraw[-,black!40!green,solid,line width = 0.9pt](1.75mm,0mm)rectangle(3mm,1.2mm);\draw[-,black!40!green,solid,line width = 0.9pt](0mm,0.6mm) --(5mm,0.6mm)}}}
\newcommand{\bluecircle}{\raisebox{0pt}{\protect\tikz{\protect\filldraw[-,blue,solid,line width = 0.9pt](2.375mm,0.6mm)circle(2pt);\protect\draw[-,blue,solid,line width = 0.9pt](0mm,0.6mm) --(5mm,0.6mm)}}}
\newcommand{\bluesquare}{\raisebox{0pt}{\tikz{\filldraw[-,blue,solid,line width = 0.9pt](1.75mm,0mm)rectangle(3mm,1.2mm);\draw[-,blue,solid,line width = 0.9pt](0mm,0.6mm) --(5mm,0.6mm)}}}
\newcommand{\greencircle}{\raisebox{0pt}{\tikz{\filldraw[-,black!40!green,solid,line width = 0.9pt](2.375mm,0.6mm)circle(2pt);\draw[-,black!40!green,solid,line width = 0.9pt](0mm,0.6mm) --(5mm,0.6mm)}}}
\newcommand{\redcircle}{\raisebox{0pt}{\tikz{\filldraw[-,red,solid,line width = 0.9pt](2.375mm,0.6mm)circle(2pt);\draw[-,red,solid,line width = 0.9pt](0mm,0.6mm) --(5mm,0.6mm)}}}
\newcommand{\redsquare}{\raisebox{0pt}{\tikz{\filldraw[-,red,solid,line width = 0.9pt](1.75mm,0mm)rectangle(3mm,1.2mm);\draw[-,red,solid,line width = 0.9pt](0mm,0.6mm) --(5mm,0.6mm)}}}
\newcommand{\orangecircle}{\raisebox{0pt}{\tikz{\filldraw[-,orange,solid,line width = 0.9pt](2.375mm,0.6mm)circle(2pt);\draw[-,orange,solid,line width = 0.9pt](0mm,0.6mm) --(5mm,0.6mm)}}}
\newcommand{\orangesquare}{\raisebox{0pt}{\tikz{\filldraw[-,orange,solid,line width = 0.9pt](1.75mm,0mm)rectangle(3mm,1.2mm);\draw[-,orange,solid,line width = 0.9pt](0mm,0.6mm) --(5mm,0.6mm)}}}

\newcommand{\magentadowntraingle}{\raisebox{0pt}{\tikz{\filldraw[-,magenta,solid,line width = 0.9pt](1.75mm,0.7mm)--(3.5mm,0.7mm)--(2.625mm,-0.8mm);\draw[-,magenta,solid,line width = 0.9pt](0mm,0.2mm) --(5mm,0.2mm)}}}
\newcommand{\magentaline}{\raisebox{2pt}{\tikz{\draw[-,magenta,solid,line width = 0.9pt](0,0) --(5mm,0);}}}
\newcommand{\bluestar}{\raisebox{0pt}{\tikz{\node[star, blue, star points=5, star point ratio=0.28, fill=blue, rotate=180, draw] at (2.5mm,0.1mm){};\draw[-,blue,solid,line width = 0.9pt](0mm,0.1mm) --(5mm,0.1mm)}}}

\newcommand{\greysquare}{\raisebox{0pt}{\tikz{\filldraw[-,sangria,solid,line width = 0.9pt](1.75mm,-0.1mm)rectangle(3.3mm,1.4mm)}}}

\begin{document}
% \pagenumbering{roman}
% % \input{titlepage.tex}
% \newpage
% %\clearpage
% \input{Declaration.tex}
% \clearpage
% \input{Aknowledgement.tex}
% \clearpage
% \input{abstract.tex}
% %\pagenumbering{roman}	
% \listoffigures
% \listoftables
% \tableofcontents
% \clearpage	
\pagenumbering{arabic}
\allowdisplaybreaks
% \chapter*{}\label{chapter:particle_wall_collison}

\begin{center}
    {\large\textbf{Review of contact models used in Discrete Element Method (DEM)}} \\[2ex]
    \textbf{Sourav Ganguli}\textsuperscript{1},
     \textbf{Partha Sarathi Goswami}\textsuperscript{2},
     \textbf{Manaswita Bose}\textsuperscript{1,*}\\[1ex]
    \textsuperscript{1,*}Department of Energy Science and Engineering, IIT Bombay, Powai, Mumbai, 400076, Maharashtra, India\\
    \textsuperscript{2}Chemical Engineering department, IIT Bombay, Powai, Mumbai, 400076, Maharashtra, India \\
    %\texttt{sourav.research.123@gmail.com}
\end{center}

\tolerance=1000
\emergencystretch=1em
\hbadness=20000
\vbadness=12000

\begin{abstract}
    This work presents a detailed review of the methods proposed to implement Mindlin's no-slip and partial slip model under constant normal loading and Mindlin Deresiewicz's extensional work on micro-slip under varying normal loading, for the simulation of granular flow. Various methods that followed Mindlin's and Mindlin and Deresiewicz's approaches for modeling the tangential contact between two spherical particles are reviewed thoroughly, first, to understand the tangential load-displacement behaviour and finally to compare the tangential coefficient of restitution obtained from each of these models with the experimentally determined values reported in the literature. The tangential load-displacement behaviour obtained from the integral method is qualitatively different from that determined by the incremental method, which accounts for the loading history. Tangential and rotational coefficient of restitution in the gross sliding regime obtained from all the models show excellent agreement with the experimental result, but in the sticking regime, the agreement is only qualitative.
\end{abstract}

\section{Introduction}\label{sec:Introduction}

In the discrete element method, the contact between the particles is modelled using a combination of either a linear or a non-linear spring, a dashpot, and a slider. The normal contact force of the latter combination is modelled by the theory proposed by Hertz and the tangential contact force is determined by the extension of the Hertz theory by \cite{R.D.Mindlin1949} and \cite{Mindlin1953}. It is important to understand the contact models and their implementation while simulating the dynamics of granular assembly using discrete element methods. The current work aims to review three different approaches that implement Mindlin's no-slip and Mindlin-Deresiewicz's micro-slip model under different loading conditions for simulations of granular flows.

\begin{figure*}[ht]
    \centering
    %\subfloat[b][]{\includegraphics[width=0.2\textwidth]{spring_dashpotsvg.eps}\label{fig:Spring_dashpot}} 
    %\subfloat[a][]{\includegraphics[width=0.55\textwidth]{particle_wall_collision_1.eps}%\label{fig:particle_wall_collision}} 
    \includegraphics[width=0.7\textwidth]{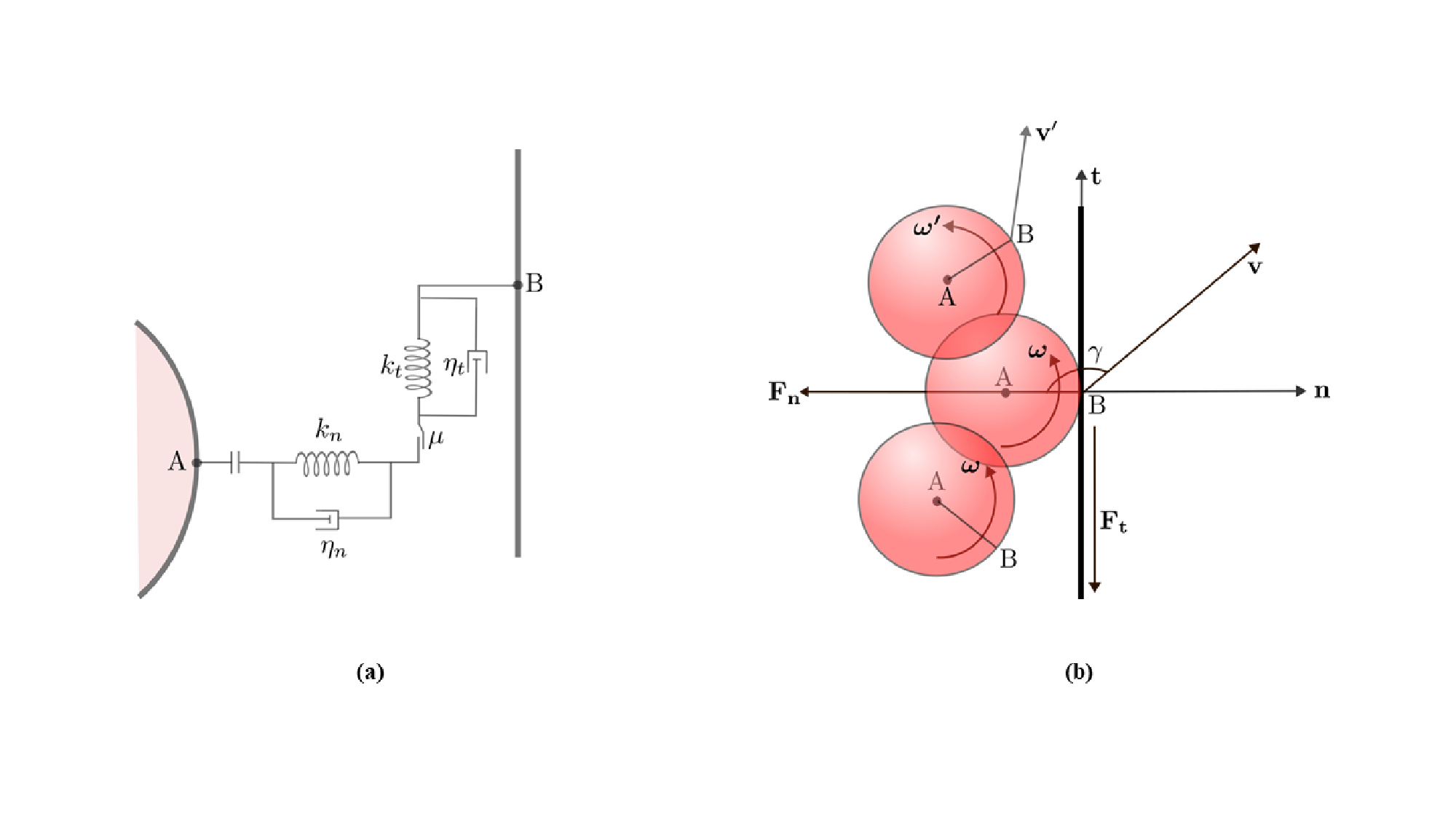}
    \caption{Schematic of (a) spring-slider-dashpot model (b)wall particle contact; $\gamma$ is the impact angle }
    \label{fig:Collision_schematic}
\end{figure*}

To begin with, we present a brief review of the literature on inter-particle contact. Particle-wall and inter-particle collision have long been a topic of interest, the very first paper in this area being authored by Sir C V Raman in 1920 \citep{Raman1920}. The research area has been active for more than a century { \citep{Vu-Quoc1999, Schwager2007, ALESHIN201214, balevivcius2018, BALEVICIUS2021685, Kosinski2020, Sinka2021, Vyas2021, DEKLERK2022,WANG2023118397, bhadra2020vibrational, Mukhopadhyay2023}} with the most recent publication on the normal coefficient of restitution appearing in 2022 \citep{Chatterjee2022}.  The reason mainly is the significant application of the discrete level methods for simulating the flow of granular material \citep{Windows-Yule2015, Syed2017, Kuang2019a, EDMANS2020618, Ikari2022, GE2023142749, YURATA2021, HORVATH2022}. 
Inter-particle or wall-particle collisions are typically modelled following two approaches, hard-sphere and soft-sphere \citep{Silbert2002, Reddy2010}. In the earlier case, the particles are assumed to be non-deformable and the post-collision velocities are completely defined by two parameters, namely, the normal ($e_n$) \citep{Schwager2007} and the tangential (rotational) ($\beta$) coefficient(s) of restitution \citep{Becker2008, Schwager2008}. On the other hand, in the case of the soft-sphere approach, the contact between the particles (or particles and wall) is modelled using a combination of spring-slider-dashpot (Figure ~\ref{fig:Collision_schematic})(a). 
% \textcolor{blue}{Figure ~\ref{fig:Collision_schematic} (b) shows the wall-particle contact and depicts the impact angle $\gamma$.}
The discrete element method is developed using the soft sphere particle contact modelling approach. The normal elastic force is modelled using a linear \citep{Cundall1979} or nonlinear spring \citep{Y.TsujiY.TsujiT.Tanaka1992, Tsuji1993, Silbert2001, Tripura2022}, the dissipation due to the inelastic nature of the collision is accounted for by the viscosity coefficient of the dashpot. The role of friction during contact is modelled based on the slip velocity at contact. It is modelled either following Coulomb friction law \citep{Brilliantov_2015}; else, the contact is modelled with the help of a spring-dashpot \citep{Tsuji1993} as shown in Figure \ref{fig:Collision_schematic}(a). In the case of non-linear springs, the normal spring constant ($k_n$) is expressed in terms of material properties such as modulus of elasticity ($Y_n$) and Poisson ratio ($\nu$), and the tangential spring constant ($k_t$) is typically selected based on the no-slip model of Mindlin \citep{johnson_1985, Kloss2011}.
\cite{Cundall1979} employed the linear spring dashpot(LSD) model to simulate the mechanical behaviour of granular assemblies comprising disks and spheres. They have used two different ratios of the tangential to the normal stiffness constants: $\frac{k_t}{k_n} = 1$ and $\frac{2}{3}$. The dissipation coefficients are considered to be proportional to the respective spring constants. The viscous dissipation coefficient is also related to the coefficient of restitution. Later, \cite{J.SchaferS.Dippel1996} derived the relationship between the viscous dissipation coefficient ($\gamma_n^l$) and properties such as the coefficient of restitution ($e_n$) and stiffness constant for LSD: ($\gamma_n^l = \frac{2 \sqrt{k_n m} \ln{e_n}}{\sqrt{\pi^2 + \ln{e_n}^2}}$). The effect of the tangential coefficient of restitution ($e_t$) on the tangential force, velocity, and slip was reported in \citep{Becker2008, Schwager2008}.

\cite{Kuwabara_1987} extended the Hertz model to account for frictional losses. In their model, the viscous force has a square root dependence on the deformation and a linear dependence on the normal velocity. They have shown that the normal coefficient of restitution depends on the one-fifth power of the approach velocity, $(1-e_n)\propto v^{1/5}$. In their analysis, they neglected the vibration in the sphere due to impact. They compared the model prediction with experimental observation of particles made of materials such as aluminum, brass, bronze, and copper. In a series of articles, P\"oschel and co-workers \citep{SchwagerPRE1998, Ramirez1999, SchwagerPRE2008, Muller2011} proposed series solutions for the equation of motion with visco-elastic contact under different conditions including a delayed recovery of shape \citep{SchwagerPRE2008}. The first term in the series in these analyses is $\propto v^{1/5}$ while the coefficients differ. Convergence of the series solution was enhanced by the Pad\'e approximation \citep{Muller2011}; however, the form of expression is difficult to implement in large-scale simulations. In addition, the explicit dependence of the normal coefficient of restitution on the approach velocity needs to be modeled only for simulations that consider a hard-sphere potential. The approach-velocity dependence of the post-contact behaviour is implicit in the Hertz model; the solution in \citep{Muller2011} is useful to estimate the values of the viscous dissipation coefficient (\cite{Sudeshna2024} and references therein). \cite{delValle2024} introduced a third-order accurate integration scheme SPIRAL for the rotational motion. A comprehensive compilation of the integration scheme used in the currently available open-source and commercial software is presented.

In a recent article, \cite{James2020} showed different numerical integration methods of Kuwabara and Kono's model; however, the model remains difficult to implement in large-scale simulations.  Contact between one spherical particle and a flat surface or two spherical particles is a classic problem with a rich history beginning with the Hertz model \citep{love1927}. Mindlin extended Hertz's solution for tangential contact forces and improved the solution in 1953 \citep{Mindlin1953}. There has been a large volume of work on different aspects of contact between two particles. \cite{ALESHIN201214} proposed a general form of solution for 2D loading using a memory diagram. A usable form of models to interpret experimental results was proposed by Balevi{\v{c}}ious and coworkers \citep{balevivcius2018, BALEVICIUS2021685}. In the present work, we limit our focus to the models that implemented Hertz-Mindlin and Hertz-Mindlin-Deresiewicz models for DEM applications.
\cite{Y.TsujiY.TsujiT.Tanaka1992} proposed to simulate inter-particle collision following the normal contact force model of Hertz and the no-slip model of Mindlin \citep{Hertz1992, Mindlin1953}. They used an arbitrary scaling factor to non-dimensionalize the linear momentum conservation equation. They integrated the normal and the tangential force based on the instantaneous position of the particle, termed the ``integral'' approach. 
The approach followed by \cite{Y.TsujiY.TsujiT.Tanaka1992} has since been quite widely used \citep{DiRenzo2005, Tripura2022} and is implemented in commercial and open-source software \citep{Kloss2011}. 

The importance of considering loading history while determining the tangential deformation has been discussed in \citep{Iwashita1985, Walton1986, Vu-Quoc1999}. \cite{Walton1986} used an expression for the effective tangential spring stiffness coefficient that approximates the theory of \cite{Mindlin1953}. \cite{Vu-Quoc1999} analyzed four different loading-unloading situations following the theory of \citep{Mindlin1953} and explained the tangential force displacement behaviour under varying normal loads. They have reviewed two different normal force-displacement (NFD) models, one directly following Hertz theory and the other based on a simplified linear elastoplastic model, in which the parameters are determined from the experimental coefficient of restitution. They performed simulations for the chute of granular matter using the tangential force-displacement model developed following Mindlin Deresiewicz's theory. They determined the post-collision properties using the results of simulations using the discrete element method.  Later, \cite{THORNTON201330} proposed a simple model combining the outcome of their simulations and the no-slip theory of \cite{R.D.Mindlin1949}, in which they have rescaled the tangential force during the unloading path, i.e., when the normal force decreases successively.

\cite{Brilliantov1995} proposed a contact model based on the stress-strain relationship of a linearly deformable material. The dissipation coefficients are considered to be arbitrary constants in their model. They modelled the friction between the colliding surfaces as a Coulomb force. \cite{Silbert2001} simulated the flow of granular particles over an inclined plane. They employed the spring-slider-dashpot model with linear and non-linear springs to simulate the contact between particles. They simulated different case studies following both approaches as proposed by \cite{Y.TsujiY.TsujiT.Tanaka1992} and \cite{Brilliantov1995} to model the viscous dissipation and compare the results. In this work, they used $\frac{k_t}{k_n} = \frac{2}{7}$, which signifies equal time-periods in the normal and tangential direction, in the case of a linear spring. Articles published at later dates (\cite{Brewster2005} and other articles)  have followed the parameters suggested by Silbert and co-workers.

In a parallel development, \cite{Maw1976} suggested an alternate method than that proposed by \cite{Mindlin1953, johnson_1985} for determining the tangential deformations. They proposed three regimes for tangential contact, namely, sticking, sticking-sliding, and gross sliding. To differentiate the regimes, they have defined a parameter:  $\psi_1 = \frac{2\left(1-\nu\right)}{\mu\left(2-\nu\right)} \frac{v_s}{v_n}$
%and $\psi_2 = -\frac{v'_s}{v_n}$,
where, $v_n$ and $v_s$ are the pre-collision normal center of mass and tangential surface velocity components of the particle relative to the wall \citep{Walton1992}, $\nu$ is the Poisson ratio and $\mu$ is the friction coefficient.  Later \cite{Walton1992} proposed a method to relate the experimentally measurable quantities, the normal and rotational coefficients of restitution ($e_n, \beta= -\frac{v'_s}{v_s}$) with the material properties such as the modulus of elasticity, Poisson ratio and the friction coefficient. They have defined a scaling for the gross sliding regime by $\frac{7}{2}(1+e_n) \mu\,\mid\cot{\gamma}\mid$. They suggested that beyond a critical value of the impact angle $\gamma_o$, the collision is in the sticking regime and $\beta $ assumes a value $\beta_o$.   
This simplified regime mapping was experimentally validated by \cite{Louge2002}; however, they commented that the simplified mapping of $\beta$ suggested by \cite{Walton1992} neglected the features shown in experiments and simulations. \cite{Louge2002} showed different regimes of contact on $\psi_1$ vs. $\psi_2$ plane \citep{maw1980} where $\psi_1 = -\frac{v_s}{v_n}$ and $\psi_2 = -\frac{v_s'}{v_n}$. Subsequently, single particle contacts were analysed on $\psi_1-\psi_2$ plane (\cite{Thornton2008, Thornton2011, DiMaio2004}, and references therein). Thornton showed that there is no unique scaling or normalizing factor for contacts in the near normal or intermediate range; however, it has been demonstrated by other researchers that for a particular set of parameters, the criteria suggested by Maw and coworkers agree well with numerical results \citep{DiRenzo2004}. In a series of articles, Di Renzo and co-workers \citep{DiMaio2004, DiRenzo2004, DiRenzo2005} investigated the effect of material properties on the force-displacement behaviour of particles at contact. They also have proposed a solution for determining the tangential (rotational) coefficient of restitution for a wide range of impact angles. A detailed review comparing the modelling approaches and the force-displacement behaviour is presented by \cite{DiRenzo2004}. The same group of researchers presented a modified model for the tangential force in 2005 \citep{DiRenzo2005}, which showed an improved agreement with the experimental results reported by \cite{Kharaz2001}. 

After a thorough review of literature, it is concluded that the methods to determine the tangential forces during contact between two spheres or a sphere and a planar surface can broadly be classified under two different categories namely, the incremental approach \citep{thornton1997, Iwashita1985}, which solves the equation of motion for particles with non-linear contact forces by determining the differential tangential force based on the incremental displacement and the integral approach which determines both tangential and normal components of the contact force based on the instantaneous position of the particles(\citep{Tsuji1993, DiMaio2004}). Table \ref{tab:diff_contact_models} summarises the notable models for determining the tangential forces. 

In a lucid explanation of Hertz-Mindlin (HM) and Hertz-Mindlin-Deresiewicz (HMD) models, \cite{Iwashita1985} have clearly articulated that an incremental approach is appropriate for accurate estimation of the tangential force-displacement behaviour. However, to the best of the authors' knowledge, an analysis of the model outcome, both, in-terms of the load-displacement behaviour and comparison of post-contact properties against the results of a single experiment, is not presented in the literature. The current work aims, first, to investigate the nature of the load-displacement behaviour obtained from the two different approaches and then to compare the rotational coefficient of restitution obtained from different methods against experimental results \citep{Kharaz2001}.  
\begin{table*}[ht!]
    % \centering
    \small
    \caption{{Summary of key literature implementing HM and HMD models for DEM simulations }}
    \begin{tabular*}{\textwidth}{p{0.13\textwidth}p{0.2\textwidth}p{0.2\textwidth}p{0.28\textwidth}}
         % \toprule       
         \hline 
         Author (year) & Normal Force & Tangential Force & Comments\\%& Contact behavior\\
         % \midrule
         \hline
         \multicolumn{4}{c}{Incremental approach for tangential force} \\% \hline
         % \midrule
         \hline
         \cite{Walton1992}  & {Integral force based on Elasto-plastic model} & Incremental tangential force based on Mindlin micro slip model & Introduced a simplified tangential contact model based on Hertz-Mindlin Deresiewicz model\\
         \cite{Vu-Quoc1999} & {Normal force based on both Hertz model and the model based on Walton et al. \cite{Walton1986}} & {Mindlin and Deresiewicz model} & Showed tangential load-displacement behaviour for different situations of increase or decrease of Normal and Tangential contact force\\
         \cite{THORNTON201330} & Integral normal force based on Tsuji et al. \cite{Y.TsujiY.TsujiT.Tanaka1992}  & Incremental tangential force with scaling & Divergence of tangential traction near the end of the contact surface is avoided in Mindlin's no-slip model \\ %\hline%& Resolves the spurious energy generation for near normal contact \\ 
         % \midrule
         \hline
         % & Integral approach for &tangential force & \\ \hline
         \multicolumn{4}{c}{Integral approach for tangential force} \\ %\hline
         % \midrule
         \hline
         \cite{Y.TsujiY.TsujiT.Tanaka1992} & Normal force based on instantaneous normal displacement considering the theory of Hertz (Integral normal force)  & Integral tangential force & Spurious energy generation is observed in certain parameter range for near normal contacts \\ %& Spurious energy generation for near normal contact\\  
         %\hline
         \cite{DiRenzo2005} & Integral normal force \cite{Y.TsujiY.TsujiT.Tanaka1992} & Scaled integral tangential force & Spurious energy generation is avoided\\ %& Resolves the spurious energy generation for near normal contact \\ 
         %\hline
         %\hline
        % ``Scaled normal force" model & Scaled integral normal force & Integral tangential force \\  %& Resolves the spurious energy generation for near normal contact \\ 
         %\hline
         % Current model B & Incremental normal force & Integral tangential force & Result is similar to Tsuji et al. (1992) \\ 
         %\hline
         %''Incremental force" model & Incremental normal force & Incremental tangential force \\ %& Resolves the spurious energy generation for near normal contact \\
         % \botrule       
         \hline
    \end{tabular*}
    \label{tab:diff_contact_models}
\end{table*}

%\section{\label{sec:Model}Model Description}

Among the models presented in literature, we have selected the incremental scaled tangential force model by Thornton and co-workers\citep{THORNTON201330}, integral Hertz-Mindlin no-slip model by Tsuji et al  \citep{Tsuji1993}, and integral scaled tangential model by Di Renzo and Di Maio \citep{DiMaio2004} for detailed analysis. The task now is first, to understand the tangential load-displacement behaviour obtained using each of the models and finally to compare the tangential coefficient of restitution obtained from these models against the experimental % The HM model \citep{Iwashita1985, johnson_1985} is then employed to reproduce the experimental 
observation reported by Kharaz et al. \cite{Kharaz2001}. To that goal, we have numerically solved the equation of motion to simulate the contact between a spherical particle and a planar wall with appropriately selected material properties \citep{Silbert2001, Reddy2010}. 

\section{Discussion on the non-linear model}\label{sec:discussion_nonlinear_model}

Before proceeding with the implementation methods and analysis, we present a very brief discussion on Mindlin \citep{R.D.Mindlin1949} and Mindlin-Deresiewicz \citep{Mindlin1953} models. The normal force between two spheres at contact, based on the Hertz model, is expressed as $F_n=K_n\delta_n^{\frac{3}{2}}$ in which $K_n = \frac{4}{3} Y_n^{*} \sqrt{R}$ is an arbitrary constant (also known as secant stiffness). The normal stiffness is defined as $k_n = 2 Y_n^* \sqrt{R \,\delta_n}$. Mindlin \citep{R.D.Mindlin1949} and Mindlin and Deresiewicz \cite{Mindlin1953} proposed a theory for determining the tangential force applied at the contact surface by one particle on the other for different conditions; for example, Mindlin proposed models for no-slip conditions under constant normal loading and defined the tangential stiffness as $k_t = 8G^* \sqrt{R\,\delta_n}$. The ratio of the tangential to the normal stiffness coefficient, defined as $\frac{4 G^{*}}{Y_n^*}$, is simplified for two contacting bodies with the same material as:
\begin{equation}
   \kappa = \frac{k_t}{k_n} = \frac{2\left(1-\nu\right)}{\left(2-\nu\right)}
    \label{eq:cntoct}
\end{equation}
where, $\nu$ is Poisson ratio, effective Young's modulus: $Y_n^* = \left[\frac{1-\nu_1^2}{Y_{n1}} +\frac{1-\nu_2^2}{Y_{n2}} \right]^{-1}$, and effective shear modulus $G^* = \left[\frac{2-\nu_1}{G_1} +  \frac{2-\nu_2}{G_2}\right]^{-1}$. $Y_n$ and $G$ are the modulus of elasticity and shear modulus of individual material, respectively. 
$R$ represents the radius of the particle, the subscripts n and t denote the normal and the tangential directions, respectively (Figure \ref{fig:Collision_schematic}b). $\delta_n$ is the normal displacement.
%, the text subscripts `n' and `t' represent the normal and tangential direction (Figure \ref{fig:Collision_schematic}b).} 
 It is evident from Eq.~\ref{eq:cntoct} that for the valid range of Poisson ratio $\left(0 < \nu \le 0.5\right)$ for a simple isotropic material, the ratio of the tangential to the normal stiffness coefficient ($\kappa=\frac{k_{t}}{k_n}$)
%($\frac{k_t}{k_n}$), is always less than 1. The ratio 
lies between 0.667 and 1 $\left[ 0.667 \le \kappa < 1\right]$. \cite{R.D.Mindlin1949} in his article, has commented that this ratio will always remain less than unity; however, it will never fall below 0.5 for no-slip condition. He also proposed a model which considers slip at the contact surface and expressed the effective tangential stiffness as $k_{t,\mathrm{eff}} = k_{t} \left[1-\frac{F_t}{\mu F_n}\right]^{\frac{1}{3}}$. He compared the tangential load obtained from the no-slip model with that considering slip at the periphery of the contact surface and showed that the tangential traction diverges at the edge of the contact surface if slip is not considered.

\cite{Mindlin1953} extended the work of \cite{R.D.Mindlin1949} for varying normal loading and further modified the effective tangential stiffness constant $\left(k_{t,\mathrm{eff}}\right) $ for various situations, namely, increasing, decreasing or oscillating tangential load under constant normal force, both normal and tangential loads increasing or decreasing and either of normal or tangential increasing while the other decreasing. They also presented analyses of different points in the loading and unloading curve. \cite{THORNTON1988}, (also in \citep{Iwashita1985}) reformulated the expression of the effective tangential stiffness as $k_{t,\mathrm{eff}} = k_t \theta \pm \mu \left(1 - \theta\right) \frac{\Delta F_n}{\Delta \delta_t}$, instead of a purely elastic form presented by Mindlin and Deresiewicz \cite{Mindlin1953}. Here $\theta$ is expressed as: 
\begin{eqnarray}
    \theta &=& \left(1 - \frac{F_t + \mu \Delta F_n}{\mu F_n}\right)^{\frac{1}{3}}\\
    \theta &=& \left(1 - \frac{F_t^* - F_t  + 2\mu \Delta F_n}{\mu F_n}\right)^{\frac{1}{3}}\\
    \theta &=& \left(1 - \frac{F_t - F_t^{**} +2 \mu \Delta F_n}{\mu F_n}\right)^{\frac{1}{3}}
\end{eqnarray}
Here $F_{t}^{*}$ and $F_{t}^{**}$ represents tangential components of force at the load reversal points \citep{Thornton2011}. 
Three equations are for three different loading regimes; Equation 2 \footnotemark is applicable for $\Delta \delta_t > 0$, and the first loading path; Equation 3 is for $\Delta \delta_t < 0$ and for the unloading path, and Equation 4 is for, the incremental tangential displacement, $\Delta \delta_t > 0$ and for reloading or second loading path.
\footnotetext{Symbol $\Delta$ represents incremental change in variables}
Thornton and coworkers, while re-expressing $\Delta \delta_t$ as $\Delta \delta_t = \frac{1}{8 G^* a}\left(\pm\mu \Delta F_n + \frac{\Delta F_t \mp \mu \Delta F_n}{\theta}\right)$, noted that if $\left|{\Delta \delta _t}\right| < \frac{\mu \Delta F_n}{k_{t}}$ for $\Delta F_n > 0$, the state point on the tangential load-displacement profile corresponding to $F_n + \Delta F_n$ will not be satisfied (Point B in Figure 7, pg 331 of \citep{Mindlin1953}). %This is because Mindlin-Deresiwicz theory is based on simple loading history. 
This problem is numerically overcome by using the initial tangential stiffness coefficient as $k_{t} = 8 G^* \sqrt{R \delta_n}$ and following an incremental approach of integrating the equation until $k_{t} \Sigma |\Delta \delta_t| > \mu \Sigma \Delta F_n$, is satisfied (see \citep{Iwashita1985}). %Hertz-Mindlin-Deresiewicz (HMD) model \citep{Mindlin1953} has been implemented in simulation of granular flows \citep{Vu-Quoc1999, Walton1986} as shown in Table ~\ref{tab:diff_contact_models}. 
In a series of articles, Thornton and coworkers \citep{Thornton2008, Thornton2011, THORNTON201330} analyzed the single particle contact and proposed a simple model based on the no-slip model of Mindlin \citep{R.D.Mindlin1949, thornton1997}, in which he suggested re-scaling of the tangential force during unloading.

\section{Determination of tangential contact force}\label{sec:Tangential_contact_force}

We now discuss the three approaches that implement the Hertz model to determine the normal contact force and the Mindlin no-slip model and the modified Mindlin-Deresiewicz model for the tangential contact between two spherical particles to simulate the flow of granular materials. We also present the tangential force-displacement plot following each of these approaches. To that end, we have solved the second law of motion (Eqs \ref{eq:linear_momentum}, \ref{eq:angular_momentum}) using the finite difference method and the velocity Verlet scheme to simulate the contact between a sphere and a flat plate.
\begin{eqnarray}
    m\frac{d\mathbf{v}}{dt} = \mathbf{F}\label{eq:linear_momentum} \\
    I{\frac{d\boldsymbol{\omega}}{dt}} = \mathbf{F} \times \mathbf{R} \label{eq:angular_momentum} \\
    \mathbf{F}=\mathbf{F_n}+\mathbf{F_t}     \label{eq:total_force}
\end{eqnarray} 
Here, m and $I$ represent the mass and the moment of inertia of the particle, $\mathbf{v}$ and $\boldsymbol{\omega}$ represents the particle linear and angular velocity, and $\mathbf{F}$ is the total, $\mathbf{F}_n$ and $\mathbf{F}_t$ are the normal and the tangential forces acting on the particle (Eq. \ref{eq:total_force}).
The numerical values used in this article are the same as those used by \cite{THORNTON201330} and those reported in experiments by \cite{Kharaz2001}. Numerical values used in the simulation are provided in the captions of relevant figures. 
%Results obtained from a Matlab script were compared with the results obtained from the open-source software LIGGGHTS for verification of the code. 
Tables ~\ref{tab:Thornton case_parameter}
and ~\ref{tab:Kharaz_case_parameter} lists the parameters used in the present work.

\begin{table}[ht]
\caption{List of parameters used in simulation for generating Figures \ref{fig:Integral_diff_model}, \ref{fig:Incremental_diff_model}, and \ref{fig:Total_Energy_nospin_en1};% and \ref{fig:Incremental_diff_sgn};
these values are same as reported in \cite{THORNTON201330}}\label{tab:Thornton case_parameter}%
\begin{tabular}{@{}lll@{}}
% \toprule
\hline
Simulation parameter & Symbol  & Value \\%& Column 4\\
% \midrule
\hline
Angle of impact ($^{\circ}$)    & $\gamma$   & 85--178 \\%& data 3  \\
Impact velocity (m/s) & \textbf{v} & 5 \\
Poisson's ratio    & $\nu$   & 0.1, 0.3  \\%& data 6  \\
Friction coefficient    & $\mu$   & 0.1, 0.4  \\%& data 9\footnotemark[2]  \\
Young's modulus (GPa) & $Y_n$ & 70 \\
Particle diameter (mm) & $\mathrm{d_{P}}$ & 50 \\
Particle density (kg/$\mathrm{m^3})$ & $\rho_\mathrm{P}$ & 2650 \\
% \botrule
\hline
\end{tabular}
\end{table}

\begin{table}[ht]
\caption{List of parameters used in simulations for generating Figures \ref{fig:Model_variation_beta_sai} and \ref{fig:diff_model_kharaz_cond}; these values are reported in \cite{Kharaz2001}}\label{tab:Kharaz_case_parameter}%
\begin{tabular}{@{}lll@{}}
% \toprule
\hline
Simulation parameter & Symbol  & Value \\%& Column 4\\
% \midrule
\hline
Angle of impact ($^{\circ}$)    & $\gamma$   & 85--178 \\%& data 3  \\
Impact velocity (m/s) & \textbf{v} & 3.9 \\
Poisson's ratio (wall)    & $\nu_1$   & 0.25  \\%& data 6  \\
Poisson's ratio (particle)    & $\nu_2$   & 0.23 \\
Friction coefficient    & $\mu$   & 0.092  \\%& data 9\footnotemark[2]  \\
Young's modulus (wall) (GPa) & $Y_{n1}$ & 70 \\
Young's modulus (particle) (GPa) & $Y_{n2}$ & 380 \\
Particle diameter (mm) & $\mathrm{d_{P}}$ & 5 \\
Particle density (kg/$\mathrm{m^3})$ & $\rho_{\mathrm{P}}$ & 4000 \\
% \botrule
\hline
\end{tabular}
\end{table}

\subsection{Incremental Tangential Force Model}\label{sec:Incremental_model_discussion} 
\cite{THORNTON201330} has proposed a simplified tangential force-displacement model in which they have used the tangential stiffness constant of the no-slip model of \cite{R.D.Mindlin1949} but rescaled the tangential force of the current time-step using the ratio of the current to the previous tangential stiffness (in other words, the square root of the ratio of the current to the previous normal displacement) during the unloading path of the normal force-displacement. They argued that the tangential force needs to be scaled down during the unloading path as the decreasing normal load leads to a reduced contact area \citep{THORNTON201330}. They have used an incremental approach (Eq ~\ref{eq:tangential_force_loading_thornton}) for determining the tangential force. 
The scaled tangential force during the decreasing normal loading path is expressed by Eq ~\ref{eq:tangential_force_unloading_thornton}. 

\vspace{4pt}
\noindent
For $\left(\Delta{\mathbf{F_{n}}> 0}\right)$,

\begin{equation}
    % \mathrm{for}& \left(\Delta{\mathbf{F_{n}}\geq 0}\right), \nonumber
    % \\
    \mathbf{F_{t}}^{i}=\mathbf{F_{t}}^{i-1}-\frac{dF_{t}^{i}}{d\delta_{t}^{i}}\left(\boldsymbol{\delta_{t}}^{i}-\boldsymbol{\delta_{t}}^{i-1}\right)    \label{eq:tangential_force_loading_thornton}
\end{equation}
for $\left(\Delta{\mathbf{F_{n}}< 0}\right)$,
    % \\
    % \mathrm{for}& \left(\Delta{\mathbf{F_{n}}< 0}\right), \nonumber
\begin{equation}
    \mathbf{F_{t}}^{i}=\mathbf{F_{t}}^{i-1}\,\frac{k_{t}^{i}}{k_{t}^{i-1}}-\frac{dF_{t}^{i}}{d\delta_{t}^{i}}\left(\boldsymbol{\delta_{t}}^{i}-\boldsymbol{\delta_{t}}^{i-1}\right)  \label{eq:tangential_force_unloading_thornton}
\end{equation}
Substituting for the tangential stiffness coefficients Eq. \ref{eq:tangential_forcedelta_unloading_thornton} is obtained;

\begin{equation}
    \mathbf{F_{t}}^{i}=\mathbf{F_{t}}^{i-1}\sqrt{\frac{\delta_{n}^{i}}{\delta_{n}^{i-1}}}-\frac{dF_{t}^{i}}{d\delta_{t}^{i}}\left(\boldsymbol{\delta_{t}}^{i}-\boldsymbol{\delta_{t}}^{i-1}\right)  \label{eq:tangential_forcedelta_unloading_thornton} 
    % \\
    % \mathrm{for}& \left(\Delta{\mathbf{F_{n}}< 0}\right) \nonumber 
    \end{equation}
    % \end{equation}
    where, 
    \begin{equation}
    \frac{dF_{t}^{i}}{d\delta_{t}^{i}}=8G^{*}\sqrt{R\delta_{n}^{i}}  \label{eq:tangential_stiffness_thornton}
\end{equation}

 Here, the superscript $i$ represents the current time step. It is to be noted that \cite{THORNTON201330} determined the normal force based on the instantaneous values of normal displacement.

\begin{figure*}[ht]
    \centering
    %\includegraphics[width=0.5\textwidth]{Ft_deltat_incre_Thorn1.jpg}
   % \subfloat[a][]{\includegraphics[width=0.45\textwidth]{Ft_deltat_incre_Thorn1.eps}\label{subfig:Incremental_diff_model_thornton_condition}} 
   % \subfloat[b][]{\includegraphics[width=0.45\textwidth]{Ft_deltat_Thornton_m04nu01.eps}\label{subfig:Incremental_diff_model_extreme_condition}} 
   \includegraphics[width=\textwidth]{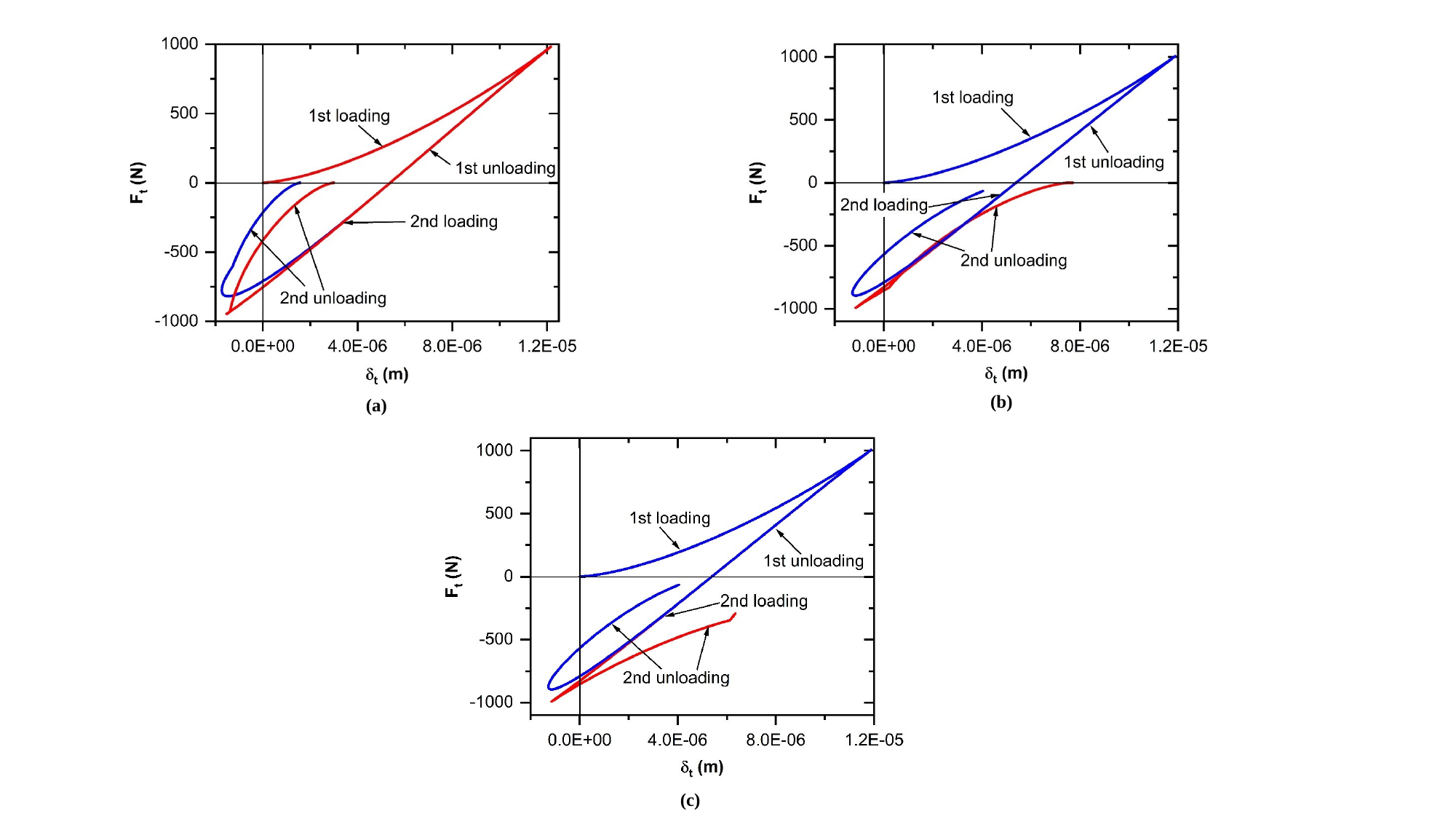}

    \caption{Plot of tangential force vs. displacement obtained using the incremental approach with the re-scaled tangential force as proposed by \cite{THORNTON201330} (\protect\blueline) and without any re-scaling of the tangential force \cite{Iwashita1985} (\protect\redline) for Poisson's ratio of (a)  0.3  \citep{THORNTON201330} and (b) 0.1, for $\mu$ of 0.1, angle of impact ($\gamma$) of $175^{\circ}$, and (c) Poisson's ratio of 0.1, $\mu=0.4$ and angle of impact ($\gamma$) of $178^{\circ}$ ; $Y_n$ of 70 GPa, particle diameter ($d_{p}$) of 50 mm and density ($\rho_{p}$) of 2650 kg/\rm{m$^{3}$} and impact velocity of 5 m/s for all cases; $F_{t}$ obtained from simulations are multiplied by -1 while plotting, to maintain consistency with the Figure A4 published in \citep{THORNTON201330}, in the current work, force is applied on the particle by the wall, in \citep{THORNTON201330} force on the wall by the particle is considered.}
    \label{fig:Incremental_diff_model}
\end{figure*}

 The incremental method to determine the normal force ($\mathbf{F}_{n}$) is expressed as equation \ref{eq:Normal_force_incremental} and \ref{eq:Normal_incremental_stiffness} \cite{Iwashita1985} and Equations \ref{eq:Tangential_force_incremental_noscaling} and \ref{eq:tangential_incremental_stiffness_noscaling} express the incremental nonscaled tangential force.
 Thornton and coworkers \citep{THORNTON201330} compared the rescaled tangential force model with the Hertz-Mindlin no-slip model without considering any scaling of tangential force (Eqs ~\ref{eq:Tangential_force_incremental_noscaling} -- ~\ref{eq:tangential_incremental_stiffness_noscaling}) and \cite{THORNTON201330} commented that the improvement due to the re-scaling was significant. It is to be noted that the normal force between elastic particles determined irrespective of the integral or incremental method agree well with each other \citep{Iwashita1985}. Figure ~\ref{fig:Incremental_diff_model}(a) shows a comparison in force-displacement plot between the scaled and the non-scaled model for the parameters reported in \citep{THORNTON201330}. The tangential load-displacement curves obtained from the two models overlap during the first loading, unloading, and part of the second loading (reloading after the first unloading) path. During the end of the second loading path, the results obtained from the two models differ. Slip at the end of the contact is $\approx$ 67\% higher if the tangential force is not re-scaled; however, the nature of the load-displacement behaviour remains qualitatively similar. 
{
\begin{eqnarray}
    \mathbf{F_n}^i&=&\mathbf{F_n}^{i-1}-\frac{dF_n^{i-1}}{d \delta_n^{i-1}}\left(\boldsymbol{\delta_n}^i-\boldsymbol{\delta_n}^{i-1}\right) \label{eq:Normal_force_incremental}\\
    \frac{dF_n^{i-1}}{d \delta_n^{i-1}}&=&2Y_{n}^{*} \sqrt{R\delta_n^{i-1}} \label{eq:Normal_incremental_stiffness}\\
   \mathbf{F_t}^i&=&\mathbf{F_t}^{i-1}-\frac{dF_t^{i-1}}{d\delta_t^{i-1}}\left(\boldsymbol{\delta_t}^i-\boldsymbol{\delta_t}^{i-1}\right) \label{eq:Tangential_force_incremental_noscaling} \\
    % \frac{dF_n^{i-1}}{d \delta_n^{i-1}}&=&2Y_{n}^{*} \sqrt{R\delta_n^{i-1}} \label{eq:Normal_incremental_stiffness}\\
    % \bf{F_t}^i&=&\mathbf{F_t}^{i-1}-\frac{dF_t^{i-1}}{d\delta_t^{i-1}}\left(\boldsymbol{\delta_t}^i-\boldsymbol{\delta_t}^{i-1}\right) \\
    % \label{eq:Tangential_force_incremental_noscaling}
   \frac{dF_t^{i-1}}{d\delta_t^{i-1}}&=&8G^{*}\sqrt{R{\delta}_n^{i-1}} \label{eq:tangential_incremental_stiffness_noscaling}
    \end{eqnarray}
 }   

Figure ~\ref{fig:Incremental_diff_model}(b) shows the same tangential load-displacement plot for another set of parameters. In this case, the Poisson ratio is 0.1, lower than the case shown in Figure ~\ref{fig:Incremental_diff_model}(a). It is observed in Figure ~\ref{fig:Incremental_diff_model}(b) that the curvature of the load-displacement plot is qualitatively different if the tangential force is not re-scaled and the slip during the second unloading path is much larger as compared to the displacement obtained from a scaled tangential incremental model of Thornton and coworker. 
%The difference in the tangential load-tangential displacement behaviour during the end of the second unloading path between the non-scaled and scaled incremental tangential force model may be because the tangential traction diverges to infinity near the edge of the contact surface if the partial slip is not considered \citep{R.D.Mindlin1949}.
As the Poisson ratio decreases, i.e., for more rigid materials, the effectiveness of the re-scaled tangential force model is apparent (Figure ~\ref{fig:Incremental_diff_model}(b)), which shows a diverging tangential displacement during the second unloading path if the force is not re-scaled. The diverging tangential diplacement during the second unloading is also observed for the third case (Figure \ref{fig:Incremental_diff_model}c) where friction coefficient is 0.4, angle of impact ($\gamma$) is $178^{\circ}$ and Poisson's ratio ($\nu$) is 0.1.

% \begin{figure}
%     \centering
%     \subfloat[a][]{\includegraphics[width=0.5\textwidth]{Ft_deltat_Thornton_m04n01.jpg}} \label{fig:Incremental_diff_model_extreme_condition}
%     %\includegraphics[width=0.5\textwidth]{Ft_deltat_Thornton_m04n01.jpg}
%     \caption{Tangential load displacement plot for (a) incremental force model (\protect \redline) and (b) Thornton et al. (2013) (\protect \blueline)}
% \end{figure}

%\begin{figure}
 %   \centering
  %  \includegraphics[width=0.5\textwidth]{hm_fig.jpeg}
   % \caption{Heatmap plot of $\Delta E_{spring}$ for second loading-unloading path for incremental method of Thornton et al. (2011) for different angle of impact ($\gamma$), coefficient of friction ($\mu$) and Poisson's ratio ($\nu$)}
   % \label{fig:Spring_energy_difference_heatmap}
%\end{figure}

\subsection{Integral Tangential Force}

\subsubsection{Non-Scaled Tangential force}
\cite{Y.TsujiY.TsujiT.Tanaka1992} implemented the Hertz-Mindlin noslip model and determined the
normal and the tangential forces as:
\begin{align}
     \mathbf{F_{n}} &= -K_n\delta_{n}^{\frac{3}{2}}\mathbf{n} - \gamma_n \dot{\delta}_n\mathbf{n} \label{eq:normal_force_integral}
     \\
    \mathbf{F_t} &= \mathrm{min}\left(\mid-k_t^c \sqrt{\delta_n}\,\boldsymbol{\delta_t} - \gamma_t \boldsymbol{\dot{\delta}_t}\mid, \mid\mu \mathbf{F_n}\mid\right)
    \mathbf{t} \label{eq:tangential_force_integral}
    \\
    K_{n}&=\frac{4}{3}Y_{n}^*\,\sqrt{R} \label{eq:Normal_sitffness_integral}
    \\
    k_t^c &= 8G^{*}\sqrt{R} %\label{eq:Tangential_stiffness_integral}
    \label{eq:Tsujiload}
\end{align}
where, $Y_n^*$ is the effective Young's modulus, defined as $\frac{1}{Y_n^{*}} = \frac{1-\nu_1^2}{Y_{n1}} + \frac{1-\nu_2^2}{Y_{n2}}$ and $\nu$ is the Poisson's ratio, and $G^*$ is the effective shear modulus. The tangential spring stiffness coefficient is $k_t = k_t^c \sqrt{\delta_n}$, $\mathbf{n}$ and $\mathbf{t}$ are unit vectors in the normal and the tangential directions. %The detailed discussion of tangential direction \textbf{t} is available in Appendix \ref{sec:tangential_direction_discussion}.  
They added a dashpot to model the dissipation due to inelasticity. They also introduced a heuristic expression to account for the displacement dependence of the dissipation coefficient and solved the momentum conservation equation to determine the coefficient of restitution as a function of the approach velocity.  

%; however they (and others following their method \cite{LIANG2023118347, WANG2021, Tripura2022, Raji2004})determined both normal and tangential force based on the instantaneous displacements.  The tangential force is obtained following Mindlin's no-slip model \citep{R.D.Mindlin1949}. They have set the tangential force $|f_t| = \mu |f_n|$ if the calculated $|f_t| \ge \mu |f_n|$. For numerical implementation, the modified displacement is determined as $\delta_t = \frac{\mu f_n}{k_{t0}}$ \citep{KRUGGELEMDEN20081523}.
 Their method of integration gives rise to two confusions.
First, the authors have termed the constant $K_n$ in Eq ~\ref{eq:Tsujiload} as the normal spring stiffness. To be precise, the $K_n$ is the secant stiffness, which is not exactly the normal spring stiffness constant understood in a typical sense as the inverse of the normal compliance. The imprecision in the definition leads to further confusion
%if $K_n$ in Eq.~\ref{eq:Tsujiload} is treated as ``stiffness'' constant, 
as the ratio of the tangential to the normal ``stiffness'' constant is expressed as
\begin{equation}
    \frac{k_t^c}{K_n} = \frac{3}{2}\times \frac{2\left(1-\nu\right)}{\left(2-\nu\right)}.
    \label{eq:stiffness_ratio_tsuji}
\end{equation}
Eq ~\ref{eq:stiffness_ratio_tsuji} indicates that the tangential stiffness is higher than the ``normal stiffness'' for valid range of Poisson ratio \citep{Tripura2022}, which is in contrast with the theory \citep{Mindlin1953, THORNTON201330}. 
Second, the integral method of integration of Hertz-Mindlin equation leads to spurious results for post-collision velocities for particles for some cases of near normal contact \citep{DiMaio2004, Thornton2011, KEISHING2020, ELATA1996229}. %Figure ~\ref{subfig:Ft_deltat_integral_thornton_aoi2nu0.01} shows the force displacement plot.

\subsubsection{Scaled Tangential Force}
 \cite{DiMaio2004} identified the shortcomings of the approach followed by Tsuji and co-workers and acknowledged the necessity of accounting for the loading history while determining the tangential force. They have corrected $k_t^{c}$ with the factor $\frac{2}{3}$. With this correction, both $K_n$ and $k_t^{c}$ are defined as secant stiffness, and the inconsistencies associated with the integral method are numerically resolved. %Flow charts showing the differences between the integral and incremental approaches are presented in Appendix \ref{sec:flowchart_model}.

 \subsubsection{Tangential Load Displacement Behaviour} 
\begin{figure*}[ht]
    \centering
    %\subfloat[a][]{\includegraphics[width=0.45\textwidth]{Ft_deltat_integral_Thorn.eps}\label{subfig:Ft_deltat_integral_thornton}}
    %\subfloat[b][]{\includegraphics[width=0.45\textwidth]{Ftdelt_inte_Thorna2n01.eps}\label{subfig:Ft_deltat_integral_thornton_aoi2nu0.01}} \\
    %\subfloat[c][]{\includegraphics[width=0.45\textwidth]{Ft_t_integral_Thornton.eps}\label{subfig:Ft_t_integral_thornton}}
    %\subfloat[d][]{\includegraphics[width=0.45\textwidth]{Ft_t_integral_extreme.eps}\label{subfig:Ft_t_integral_extreme}}
    \includegraphics[width=\textwidth]{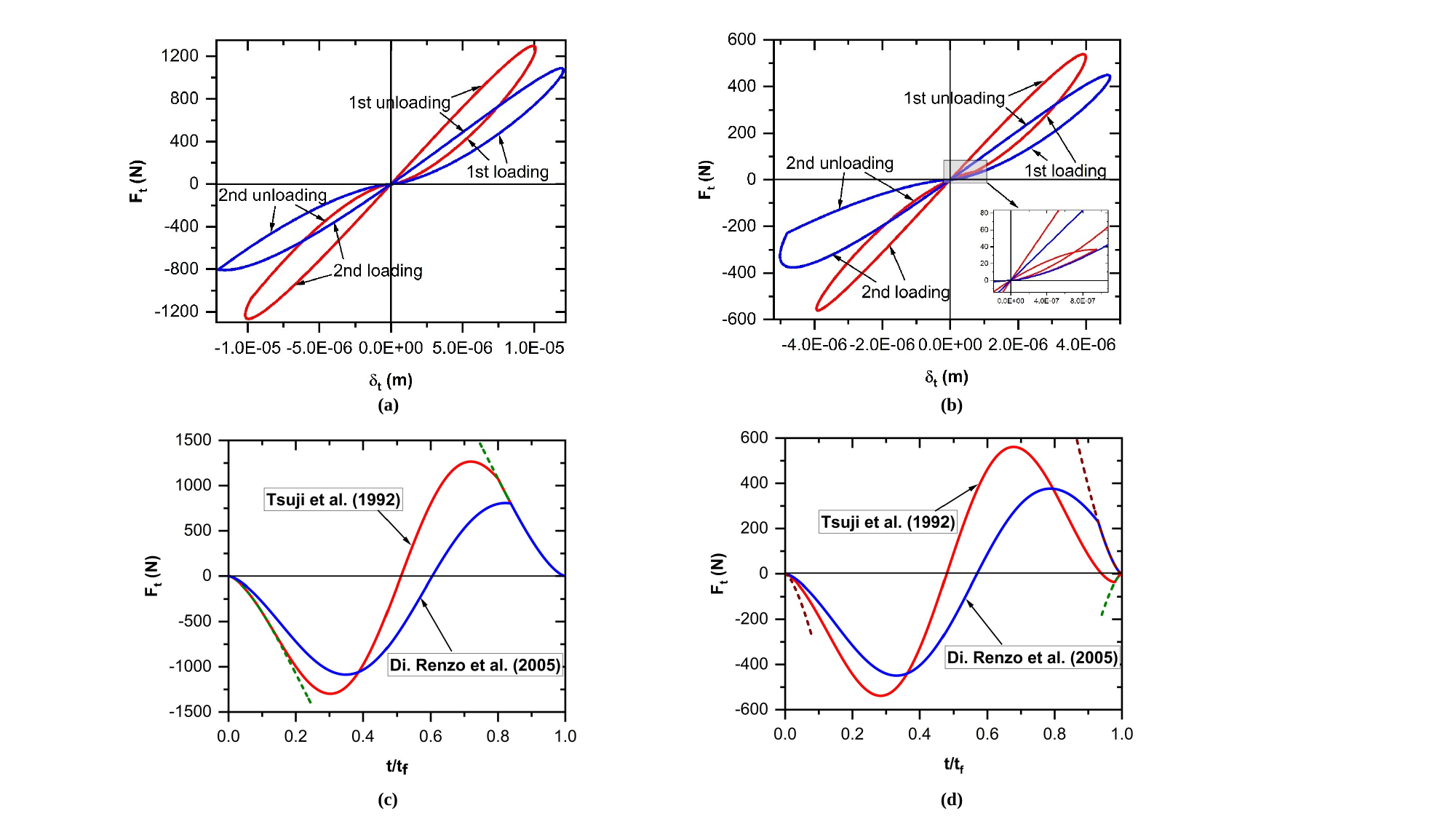}
    \caption{Tangential force vs displacement (a \& b) and Tangential force vs. normalized contact time (c \& d) determined using the integral method of Tsuji et al. (\protect\redline) and integral scaled tangential force model of Di. Renzo et al. \citep{DiRenzo2005} (\protect\blueline) for (a) \& (c) $\nu=0.3, \gamma = 175^{\circ}$, $\mu = 0.1$ and (b) \& (d) $\nu=0.1, \gamma = 178^{\circ}$, $\mu = 0.4$: $Y_n$ = 70 GPa; $\gamma_{t}=0$, particle diameter ($d_{p}$) of 50 mm and density ($\rho_{p}$) of 2650 kg/\rm{$m^{3}$} and impact velocity of 5 m/s for both cases; $F_{t}$ in (a) and (b) are multiplied by -1 while plotting, to maintain consistency with Figures 2a and 2b.}
    \label{fig:Integral_diff_model}
\end{figure*}

 The load-displacement curves obtained using the integral method, with \citep{DiMaio2004} and without \citep{Tsuji1993} scaling the tangential force are shown in Figure \ref{fig:Integral_diff_model} for two sets of parameters used before. Figures ~\ref{fig:Integral_diff_model}(a) and ~\ref{fig:Integral_diff_model}(c) show the tangential force vs displacement and time respectively for the parameters reported by  \cite{THORNTON201330} and Figures ~\ref{fig:Integral_diff_model}(b) and ~\ref{fig:Integral_diff_model}(d) show the similar plots for a material with Poisson ratio $\nu =0.1$ and $\gamma = 178 ^\circ$.
Both sets of parameters correspond to near-normal contacts. In case of the first set of parameters,  the load-displacement behaviour (Figure ~\ref{fig:Integral_diff_model}(a)) shows that the tangential force obtained from the non-scaled model is larger than that determined from the scaled model. It is also observed that for both scaled and non-scaled model, there is a certain amount of stored energy at the end of the first loading-unloading path. It is to be mentioned that there is a tangential slip at the beginning of the contact, which should lead to energy dissipation due to friction; however, the first load-displacement curve does not reflect the energy dissipation. At the end of the contact, a part of the stored energy is dissipated due to frictional slip before the particle is detached from the surface (Figure ~\ref{fig:Integral_diff_model}(c)). It is worth noting that the load-displacement behaviour obtained from the integral models is qualitatively different from that shown in Figure ~\ref{fig:Incremental_diff_model}. The tangential load-displacement plot determined using the incremental method shows a net energy dissipation at the end of the first loading-unloading path, which the integral approach cannot capture.

 Figure ~\ref{fig:Integral_diff_model}(b) shows the tangential load displacement behaviour of near normal contact between a particle with very low Poisson ratio ($\nu = 0.1$). All other parameters are kept same as the previous case. Tangential load displacement behaviour of the sphere with $\nu = 0.1$ is very similar to the one with $\nu=0.3$ (Figure ~\ref{fig:Integral_diff_model}(a) and ~\ref{fig:Integral_diff_model}(b)).  However, the inset of the figure shows that for the non-scaled model, the particle goes through a third loading-unloading path before it gets detached from the surface. It is more clearly evidenced in Figure ~\ref{fig:Integral_diff_model}(d), which shows the tangential force as a function of contact time. The curve obtained from the non-scaled integral method shows a prolonged second unloading path followed by a short third loading path before the particle finally slips and detaches from the surface. During the entire contact, the tangential force is more in case of the non-scaled model than the scaled model. This loading-unloading path leads to an overall anomalous contact behaviour. The post-contact slip velocity ($\mathbf{v_s}' = \mathbf{v_t}'+\bf{R}\times \boldsymbol{\omega}'$) has a larger magnitude than the pre-contact slip velocity ($\mathbf{v_s} = \mathbf{v_t}+\bf{R}\times \boldsymbol{\omega}$) but both have same direction resulting in $\beta = -1.05$. The inconsistency is caused because of ignoring the loading history while integrating the equation of motion in the tangential direction. However, this problem was resolved by scaling the tangential stiffness constant, \citep{DiMaio2004}, as shown by the darker (blue) curve in Figure ~\ref{fig:Integral_diff_model}(d).

 \subsubsection{Comment on energy conservation during contact:} 
\begin{figure}[ht]
    \centering
    \includegraphics[width=0.45\textwidth]{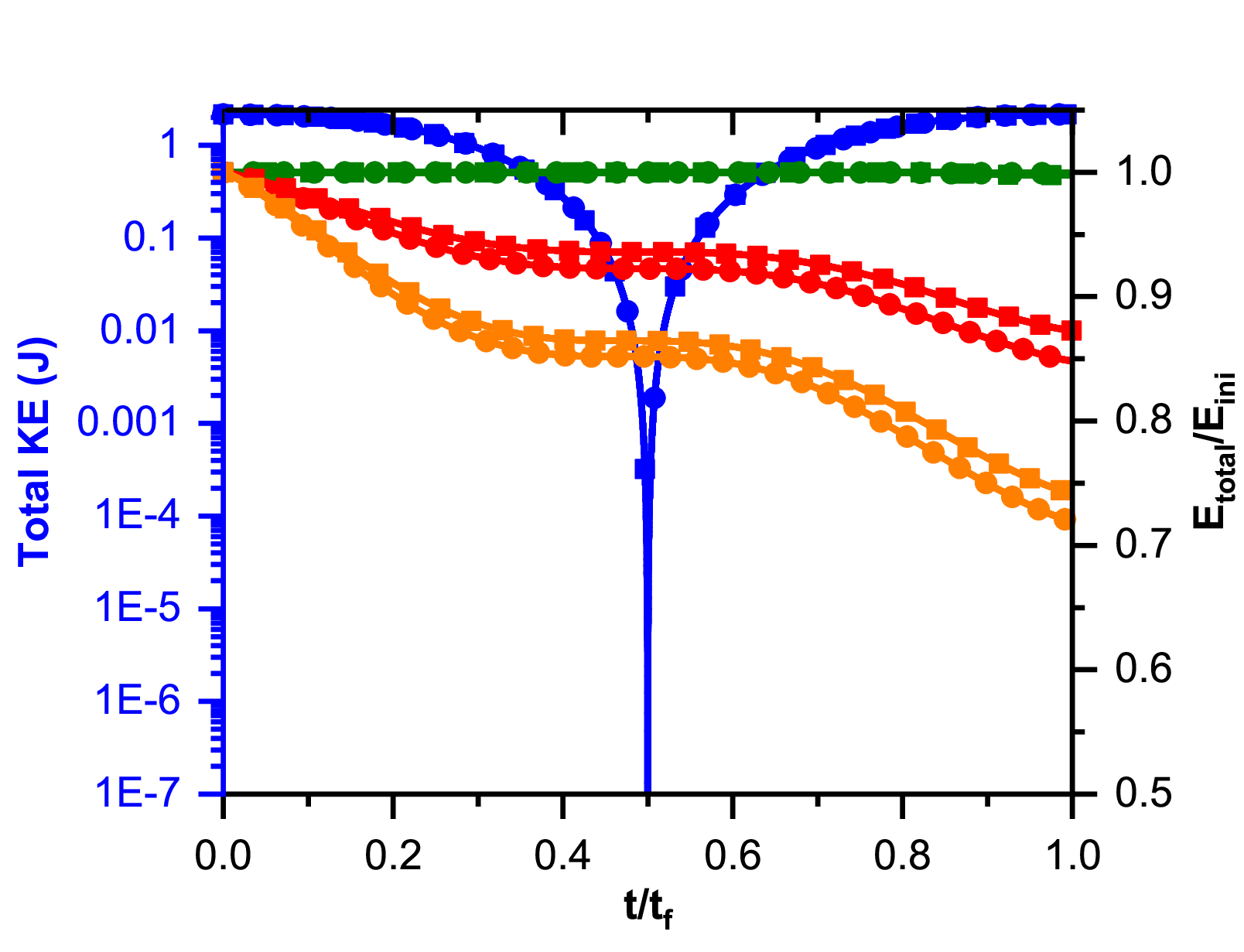} \label{fig:Total_KE_178_omega0}
    %\subfloat[b][]{\includegraphics[width=0.5\textwidth]{E_total_178_ome100.png} \label{fig:Total_KE_178_omega100}}
%    \captionsetup{justification=raggedright}
    %\caption{Variation of the total kinetic energy of the particle during contact for current model (\redline) and the model A \citep{Y.TsujiY.TsujiT.Tanaka1992} (\bluedash) for no particle initial angular velocity} %($\bm{\omega}$) and (b) initial angular velocity ($\bm{\omega}$) of 100 rad/s (Anti-clockwise)}
    \caption{Variation of total kinetic energy with time obtained from \cite{DiRenzo2005} (\protect\bluesquare) and \cite{Y.TsujiY.TsujiT.Tanaka1992} (\bluecircle) and total energy of the particle $E_{\mathrm{total}}/E_{\mathrm{ini}}$ for \cite{DiRenzo2005} (\protect\greensquare) and \cite{Y.TsujiY.TsujiT.Tanaka1992} (\protect\greencircle) for $e_{n}=1$, $E_{\mathrm{total}}/E_{\mathrm{ini}}$ for \cite{DiRenzo2005} (\protect\redsquare) and Tsuji et al. (\protect\redcircle) for $e_{n}=0.9$ and $E_{\mathrm{total}}/E_{\mathrm{ini}}$ for Di. Renzo et al. \citep{DiRenzo2005} (\protect\orangesquare) and Tsuji et al. (\protect\orangecircle) for $e_{n}=0.8$. }
    \label{fig:Total_Energy_nospin_en1}
\end{figure}

 We next perform an order-of-magnitude analysis of the spurious energy generation observed in the previous section. The kinetic energy due to the slip velocity of the contact point, $(\frac{1}{2} m v_t^2 + \frac{1}{2} I\omega^2)$ at the end of the contact is found to be $\approx$ $3$ \% higher than the pre-collision value. The term $\frac{1}{2}mv_n^2$, which is $10^3$ times higher than $(\frac{1}{2} m v_t^2 + \frac{1}{2} I\omega^2)$, remains unchanged during contact as the viscous dissipation is neglected for the case presented in Figure ~\ref{fig:Integral_diff_model}. Figure ~\ref{fig:Total_Energy_nospin_en1} shows the variation of the total kinetic energy obtained from scaled tangential and non-scaled integral methods. The difference between the results obtained from the two methods are not noticeable on the scale of the plot. Figure ~\ref{fig:Total_Energy_nospin_en1} also plots the total energy of the particle at contact ($E_{total}$) scaled by the initial energy ($E_{ini}$), $\frac{\frac{1}{2}m v_{n}^{2}+\frac{1}{2}m v_{t}^{2}+\frac{1}{2}I \omega^{2}+\int_{0}^{\delta_{n}} K_n \delta_{n}^{3/2}d\delta_{n}+\int_{0}^{\delta_{t}}\frac{1}{2}{k_t}^{c} \sqrt{\delta_{n}}\delta_{t}d\delta_{t}}{\frac{1}{2}m {v_n}_{ini}^2 + \frac{1}{2} m {v_t}_{ini}^2 + \frac{1}{2} I \omega_{ini}^2}$. The subscript ``ini" refers to the initial kinetic energy of the particle. It is observed that the total energy remains constant (within numerical accuracy of $10^{-7}$) during contact when the inelastic effect during normal contact is neglected. 
 %\textcolor{blue}{The term $\frac{1}{2}m v_{n}^2$ is $\approx 6 \times 10^2$ times higher than the term $(\frac{1}{2} m v_t^2 + \frac{1}{2} I\omega^2)$, causing total kinetic energy remains unchanged though there is an in increase of $(\frac{1}{2} m v_t^2 + \frac{1}{2} I\omega^2)$.} 
The figure also shows the ratio of the total to the initial energy of the particle during contact for particles with $e_n = 0.8, 0.9$. In these cases, the total energy of the particle at the end of contact is less than the initial energy, as expected. The reason why the anomalous behaviour of $\beta$ being less than $-1$, during the near normal contact, may have remained largely unnoticed, was mostly because simulations are typically performed for inelastic particles, for which the post-collision energy is always less than that of the pre-collision. 

\section{Comparison with Experiment: Case Study}\label{subsection:Case_study}

After discussing the load-displacement behaviour for the near normal contacts, we finally compare {all} %the two 
models against the experimental results reported by \citep{Kharaz2001}. They investigated the collision of a spherical aluminum oxide particle  (properties listed in Table~\ref{tab:Kharaz_case_parameter}) on a horizontal flat glass anvil (Table ~\ref{tab:Kharaz_case_parameter}) for different angles of impact. The coefficient of friction $\mu$ for the contacting surface is reported to be 0.092. The authors have measured the pre- and post-contact velocities and reported the tangential coefficient of restitution $e_{t}=\frac{\bf{v_{t}} ' \cdot \bf{t}}{\bf{v_{t}}\cdot \bf{t}}$, based on the tangential component of the center of mass velocity ($v_t$), for different angles of impact.  
\begin{figure*}[ht]
\centering
   % \subfloat[a][]{\includegraphics[width=0.4\textwidth]{e_t_diff_model.eps}
   % \label{fig:et_diff_model}}
    %\subfloat[b][]{\includegraphics[width=0.4\textwidth]{beta_kn_diff_model.eps}
   % \label{fig:beta_different_model}}
    %\end{center}
    %\subfloat[c][]{\includegraphics[width=0.5\textwidth]{sai_kn_diff_model.png}\label{fig:sai_different_model}}
    \includegraphics[width=0.85\textwidth]{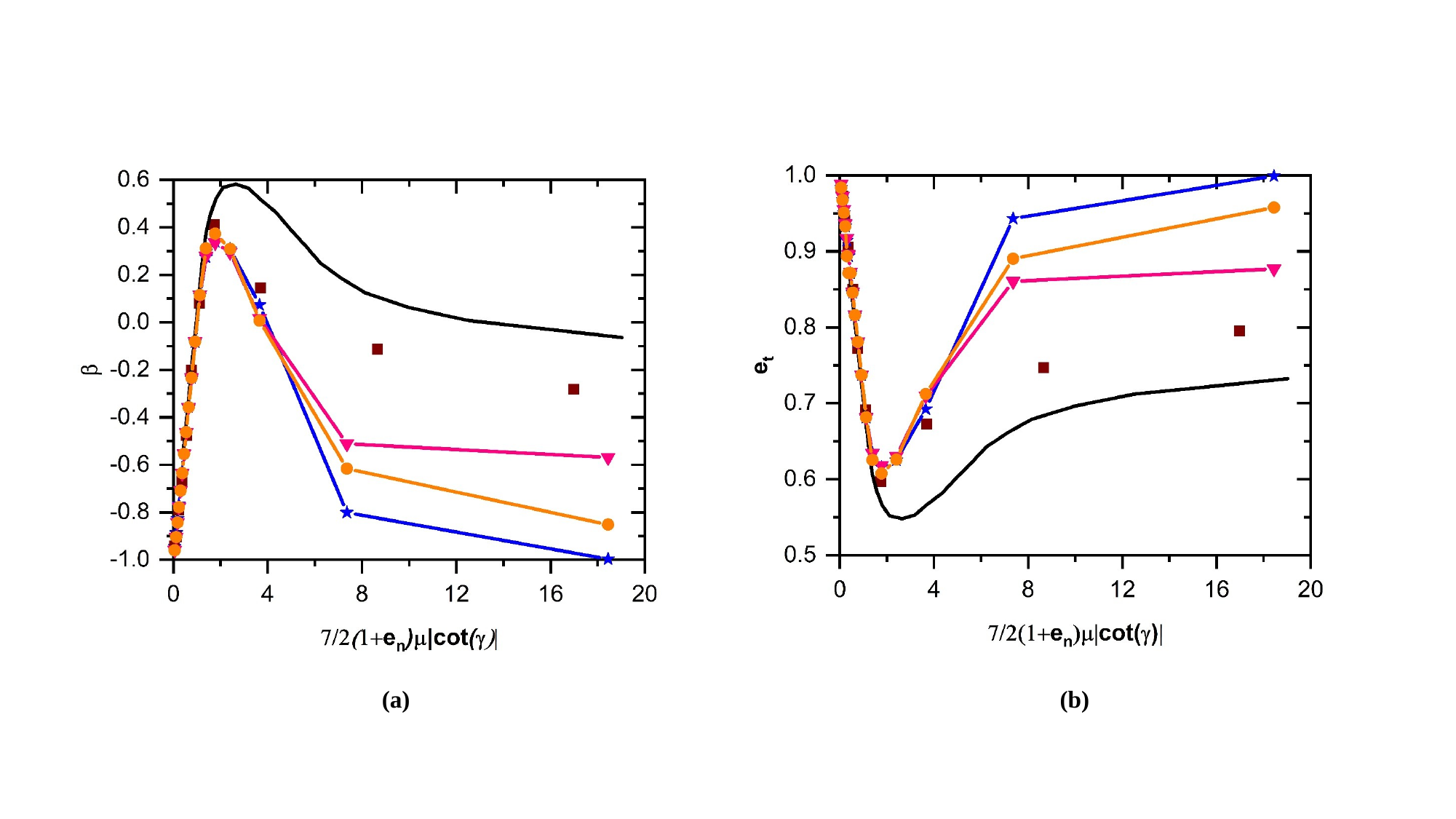}
    %\captionsetup{justification=raggedright}
    \caption{Comparison of (a) $e_{t}$ and (b) $\beta$ vs. $7/2\left(1+e_{n}\right)\mu \mid \cot\left(\gamma\right)\mid$ between the experimental results (\protect\greysquare) \citep{Kharaz2001} and simulation results based on model of \cite{DiRenzo2005} for  (\protect\blackline), Model of Tsuji et al. (1992) (\protect\bluestar), incremental force model \cite{Iwashita1985} (\protect\orangecircle) \cite{THORNTON201330}(\protect\magentadowntraingle); $\nu_{1} = 0.25$, $\nu_{2} = 0.23$, $Y_{n1} = 70$ GPa and $Y_{n2} = 380$ GPa, $\mu = 0.092$; $d_p$ = 5 mm, particle density ($\rho_p$)=4000 kg/\rm{$m^3$} and impact velocity of 3.9 m/s } 
    \label{fig:Model_variation_beta_sai}
\end{figure*}

\begin{figure*}[ht]
   \centering
       % \subfloat[a][]{\includegraphics[width=0.4\textwidth]{Ft_deltat_diff_approach.eps}\label{fig:Ft_deltat_diff_approach}}
       % \subfloat[b][]{\includegraphics[width=0.4\textwidth]{vs_t_diff_approach.eps} \label{fig:vs_t_diff_approach}}
    %\end{center}
    \includegraphics[width=0.85\textwidth]{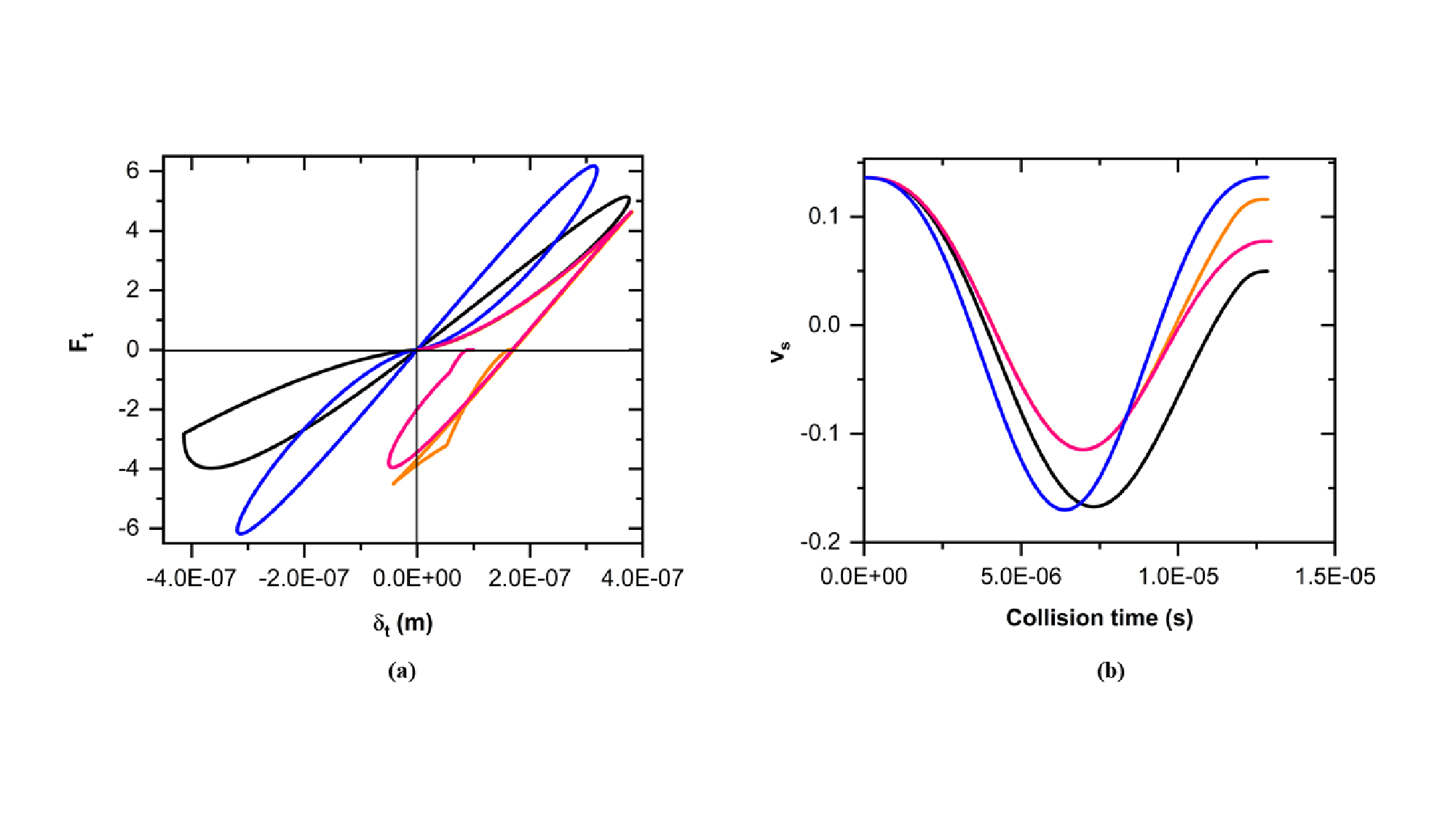}
    \caption{Variation of (a) $F_{t}$ with tangential displacement ($\delta_{t}$) and (b) tangential slip velocity ($v_{s}$) with collision time for incremental force model \cite{Iwashita1985} (\protect\orangeline), \cite{THORNTON201330} (\protect\magentaline), model of Tsuji et al. (\protect\blueline) and model of Di. Renzo et al. (\protect\blackline) for $\nu_{1}=0.25$, $\nu_{2}=0.23$ according to the experimental condition of \cite{Kharaz2001}.}
    \label{fig:diff_model_kharaz_cond}
\end{figure*}
Figure ~\ref{fig:Model_variation_beta_sai}(a) and ~\ref{fig:Model_variation_beta_sai}(b) show the plots of tangential and rotational coefficients of restitution vs $\frac{7}{2} (1+e_n)\mu\,\mid\cot{\gamma}\mid$), respectively. To summarise the observation:
\begin{enumerate}
    \item Results from all models coincide in the gross sliding regime; however, they vary quite significantly in the sticking and sticking-sliding regime. 
    \item The scaled tangential force model of Di Renzo and Di Maio under-predicts the tangential coefficient of restitution and over-predicts the rotational coefficient restitution in the sticking and the sticking sliding regime; however, their model predicts qualitative behaviour with consistency. 
    \item Coefficients of restitution, both tangential and rotational, predicted by the incremental method are in excellent agreement in the sticking sliding regime. These models over-predict the tangential coefficient of restitution and under-predict the rotational coefficient of restitution for near-normal contacts. The scaled tangential incremental method proposed by Thornton and coworkers \citep{Thornton2011} has a closer agreement with the experiment than the non-scaled incremental model. 
    \item The tangential and the rotational coefficients of restitution determined by the integral model of \cite{Tsuji1993} have excellent agreement with experimental results and with the predictions of the incremental models in the sticking-sliding regime. For the near-normal contact, the values for the coefficient of restitution obtained from the integral model are closer to the incremental models than those determined from the scaled integral model. This is counterintuitive. 
\end{enumerate}

Figure ~\ref{fig:diff_model_kharaz_cond}(a) shows the force-displacement plot obtained using the scaled and non-scaled integral, incremental and scaled tangential incremental method for a near normal contact for the collision of a 5 mm diameter spherical particle with a planer wall (properties listed in Table ~\ref{tab:Kharaz_case_parameter}) for friction coefficient  ($\mu$) of 0.092. The tangential force-displacement behaviour obtained from the incremental method is qualitatively different from that determined using the integral methods. The incremental method shows a larger amount of slip at the end of the contact. The slip velocity determined by the integral method is within $\approx 43 \% $ of that obtained using the incremental method. The interplay between the tangential force, displacement and slip velocity is unclear to us and requires in depth understanding of  Mindlin's work \citep{R.D.Mindlin1949}.

% \begin{table}[ht]
% \caption{List of symbols}\label{tab:Nomenclature}%
% \begin{tabular}{@{}ll@{}}
% \toprule
% Simulation parameter & Symbol \\
% \midrule
% Normal contact force & $F_n$ \\
% Tangential contact force & $F_t$ \\
% Normal displacement & $\delta_{n}$ \\
% Tangential displacement & $\delta_{t}$ \\
% Normal velocity & $v_{n}$ \\
% Tangential velocity & $v_{t}$ \\
% Tangential slip velocity & $v_{s}$ \\
% Angular velocity & $\omega$ \\
% Modulus of rigidity  & G \\
% Normal coefficient of restitution & $e_n$ \\
% Tangential coefficient of restitution & $e_{t}$ \\
% Rotational coefficient of restitution & $\beta$ \\
% Normal viscous dissipation coefficient & $\gamma_n$ \\
% Tangential viscous dissipation coefficient & $\gamma_t$\\
% Moment of inertia & I\\
% Particle radius & R \\
% \botrule
% \end{tabular}
% \footnotetext{Bold symbol is used for vector notation}
% \end{table}

\section{Conclusion}\label{sec:Conclusion}

To summarise, 
\begin{enumerate}
\item In this work, we have revisited the implementation of the Hertz-Mindlin and Deresiewicz model. We have examined four different existing methods in the literature \hspace{10pt} \citep{Y.TsujiY.TsujiT.Tanaka1992, DiMaio2004, Iwashita1985, Thornton2011}. 
\item Comparison against reported experimental results \citep{Kharaz2001} shows that all models agree with the gross sliding regime of contact.
\item For the near normal collision, the integral approach has a closer agreement with reported experimental results \citep{Kharaz2001} only if the tangential force is modified. The incremental approach \citep{Iwashita1985} shows a similar qualitative trend but better quantitative agreement with experimental results than the integral \citep{Tsuji1993}. The scaled tangential incremental approach \cite{THORNTON201330}, has a better quantitative agreement with the experimental results than the incremental approach.
\item Closer agreement between the integral method by Tsuji et al and the incremental methods for near-normal contact is counterintuitive. The interplay between the tangential force, displacement, and the slip velocity of the contact point is not clearly understood.
\item Based on the analysis, the scaled tangential model in \citep{THORNTON201330} appears to be the most suited method among the four reviewed for large-scale simulations. Comparison of the results based on flow of granular assembly remains as a future scope of this work.%More detailed experiments and thorough understanding of Mindlin \cite{R.D.Mindlin1949} and Mindlin and Deresiewicz \cite{Mindlin1953} model remain as future scope of work
\end{enumerate}

\section{Acknowledgment}
One of the authors, MB is grateful to Prof. K Kesava Rao for insightful discussion.

\bibliographystyle {elsarticle-harv}
  %\section*{References}
  \renewcommand{\bibname}{References}  %\renewcommand{\bibname}{}
  \bibliography{reference}

\end{document}